\documentclass[fleqn,usenatbib,useAMS]{mnras}

\usepackage{graphicx}	% Including figure files
\usepackage{amsmath}	% Advanced maths commands
\usepackage{amssymb}	% Extra maths symbols
\usepackage{multicol}        % Multi-column entries in tables
\usepackage{bm}		% Bold maths symbols, including upright Greek
\usepackage{pdflscape}	% Landscape pages

\usepackage[T1]{fontenc}
\usepackage{ae,aecompl}
\usepackage{newtxtext,newtxmath}

\title[Streaming instability: unstable epicycles]{Channels for streaming instability in dusty discs}

\author[Jaupart \& Laibe]{Etienne Jaupart$^{1}$\thanks{E-mail: \href{mailto:etienne.jaupart@ens-lyon.fr}{mailto:etienne.jaupart@ens-lyon.fr}}, Guillaume Laibe$^{1}$  \\
$^{1}$Univ Lyon, Univ Lyon1, Ens de Lyon, CNRS, Centre de Recherche Astrophysique de Lyon UMR5574, F-69230, Saint-Genis-Laval, France
}

\date{2019}

\pubyear{2019}

\begin{document}
\label{firstpage}
\pagerange{\pageref{firstpage}--\pageref{lastpage}}
\maketitle

\begin{abstract}
Streaming instability is a privileged channel to bridge the gap between collisional growth of dust grains and planetesimal formation triggered by gravity. This instability is thought to develop through its secular mode, which is long-time growing and may not develop easily in real discs. We address this point by revisiting its perturbation analysis. A third-order expansion with respect to the Stokes number reveals important features over-looked so far. The secular mode can be stable. Epicycles can be unstable, more resistant to viscosity and are identified by Green's function analysis as promising channels for planetesimals formation.

\end{abstract}


%\begingroup
%\let\clearpage\relax
%\tableofcontents
%\endgroup
\newpage

%=====================================
\section{Introduction}

Spatially resolved observations have revealed the presence of sub-structures in discs around young stars (e.g. \citealt{vanderMarel2013,Benisty2015,HLTau2015,Avenhaus2018,Andrews2018}). Whether these structures are created by planets or not is still a matter of ardent discussions. Recent direct imaging of massive planets inside the disc around PDS 70 \citep{Keppler2018,Christiaens2019,Keppler2019}, or analysis of gas kinematics \citep{Pinte2018,Teague2018,Pinte2019} suggest that at least some of these structures are indeed created by young planets. This raises the question of forming these objects in less than a typical million years. This leaves a critically short time for the solid material arising from the dusty interstellar medium to grow over $\sim$ 30 orders of magnitude in mass \citep{Chiang2010,Testi2014}. Hit-and stick collisions form millimetre pebbles relatively easily, but becomes inefficient to overcome the metre-size barrier (e.g. \citealt{Blum2008}). It has therefore been proposed that dust particles should concentrate through hydrodynamical processes in dust-rich clouds, up to the stage where gravity takes over and forms planetesimals. Proceeding to this concentration is best explained by the so-called streaming instability, which has been discovered by \citet{Youdin2005} following an idea of \citet{Goodman2000}. In thin cold discs, dust and gas exchange angular momentum through drag and drift radially with respect to each other. However, interactions between these two streams can destabilise the flows for small perturbations. Gas is then expelled in the vertical direction, leading to a local enrichment in dust \citep{Youdin2005,Youdin2007,Jacquet2011}. This behaviour is generic to a more general class of instability called resonant drag instabilities (e.g. \citealt{Squire2018,Hopkins2018}). Numerical simulations have shown that when reaching the non-linear stage, streaming instability gives rise to very high local solid concentration (e.g. \citealt{Johansen2007,JS2007,Balsara2009,Johansen2009,Tilley2010,Bai2010,Bai2010b,Bai2010c,Johansen2012,Lyra2013,Kowalik2013}) and as such is one of the corner stones of planet formation (e.g. \citealt{Dra2014,Yang2014,Simon2016,Schafer2017,Carrera2017,Schoonenberg2017}). 

So far, the instability has been mostly thought to develop through its secular mode. However, this mode has been found to grow slowly, rising concerns regarding the ability of the instability to occur in real discs. In particular, streaming instability is not thought to resist viscous damping even in moderately viscous discs ($\alpha \gtrsim 10^{-5} - 10^{-4}$, \citealt{Youdin2005}), except maybe in local pressure maxima \citep{Auffinger2018}. Streaming instability may also be quenched when the dust distribution is not monodisperse \citep{Krapp2019}.  Hence the necessity of looking for possible alternative channels. One possibility is the so-called settling instability that may develop faster \citep{Squire2018}. Another possibility has actually been suggested in the original article of \citet{Youdin2005}. They note that epicycles can become unstable but did not quantify the conditions under which this occurs. Since alternative unstable modes have not attracted much interest so far, we investigate the possible existence of complementary channels to concentrate dust. Such a mode should have a growth rate that competes with the secular mode, resists viscous damping and be favourably excited in real discs. To identify it, we revisit the perturbation analysis by obtaining an excellent approximation of the dispersion relation that factorises the epicycles and the secular mode. This study is hence organised as follows: the linear set of equations governing the evolution of a small local perturbation inside the dusty disc is presented in Sect.~\ref{sec:model_description}. The analytic study of the unstable modes is performed and stability conditions are derived in Sect.~\ref{sec:unstable}. In Sect.~\ref{sec:discuss}, we discuss the resilience against viscous damping and characterise the development of the streaming instability in real discs by the mean of Green's function analysis.

\newpage
%=====================================
\section{Equations of motion} 
\label{sec:model_description}

%-------------------------------------
\subsection{Mass and momentum conservation}
We consider a non-magnetic non self-graviting vertically isothermal inviscid  and unstratified disc orbiting a point-like central star. Dust grains are modelled by compact homogeneous spheres. Dust is treated as a continuous pressureless and inviscid phase \citep{Saffman1962}. We neglect grain growth and fragmentation. Dust and gas exchange momentum via a drag term, whose characteristic time is called the stopping time $t_{\rm stop}$. Mass and momentum conservation for gas and dust are given in the usual cylindrical coordinates by
\begin{eqnarray}
\frac{\partial \rho_\mathrm{g}}{\partial t} +  \bm{\nabla} \cdot \left(\rho_\mathrm{g} \bm{V}_\mathrm{g} \right) & = & 0 \label{eq:continuity_gas}, \\
\frac{\partial \rho_\mathrm{p}}{\partial t} +  \bm{\nabla} \cdot \left(\rho_\mathrm{p} \bm{V}_\mathrm{p} \right)  & = & 0, \label{eq:continuity_dust} \\
\frac{\partial \bm{V}_\mathrm{g}}{\partial t} + \left(\bm{V}_\mathrm{g} \cdot \bm{\nabla} \right)\bm{V}_\mathrm{g}  & = & - \Omega_\mathrm{K}^2 \bm{r} - \frac{1}{\rho_\mathrm{g}} \bm{\nabla} P + \frac{\rho_\mathrm{p}}{\rho_\mathrm{g}}\frac{\bm{V}_\mathrm{p} - \bm{V}_\mathrm{g}}{t_\mathrm{stop}} \label{eq:momentum_gas}, \\
\frac{\partial \bm{V}_\mathrm{p}}{\partial t} + \left(\bm{V}_\mathrm{p} \cdot \bm{\nabla} \right)\bm{V}_\mathrm{p}   & = &- \Omega_\mathrm{K}^2 \bm{r} - \frac{\bm{V}_\mathrm{p} - \bm{V}_\mathrm{g}}{t_\mathrm{stop}} \label{eq:momentum_dust}, 
\end{eqnarray}
where $\rho_\mathrm{g}$ and $\rho_\mathrm{p}$ denote the gas and the dust densities, $\bm{V}_\mathrm{g}$ and $\bm{V}_\mathrm{p}$ denote the gas and dust velocities, $ \Omega_\mathrm{K}$ is the orbital frequency at a given distance $r$ and $P$ is the pressure of the gas. The notations of \citet{Youdin2005} are adopted for sake of clarity. This system of equations can be either closed with an equation of state or an incompressibility condition for the gas (Boussinesq approximation). 

We follow \citet{Youdin2005,Jacquet2011} by adopting a single fluid description of the dust/gas for performing the linear stability analysis. We introduce the total density $\rho = \rho_\mathrm{g} + \rho_\mathrm{p}$ and the centre-of-mass velocity $\rho \bm{V} =  \rho_\mathrm{g} \bm{V}_\mathrm{g} + \rho_\mathrm{p} \bm{V}_\mathrm{p} $. The differential dynamics of the mixture is then unambiguously described in terms of the drift velocity $\Delta \bm{V} = \bm{V}_\mathrm{p} - \bm{V}_\mathrm{g}$ and the mass fractions $f_\mathrm{p,g} = \rho_\mathrm{p,g}/\rho$ (e.g. \citealt{Laibe2014,Lebreuilly2019}). Similarly to \citet{Youdin2005}, we close the system of equations with an incompressibility condition for practical tractability. Finally, we write the equations of motion in a the frame rotating at frequency $ \bm{\Omega}_{\mathrm{K},0} \equiv  \bm{\Omega}_{\mathrm{K}}(r_0)$ where $r_0$ is an arbitrary radius of interest. Under these assumptions, Eqs.~\ref{eq:continuity_gas} -- \ref{eq:momentum_dust} reduce to
\begin{eqnarray}
\frac{\partial \rho}{\partial t} +  \bm{\nabla} \cdot \left(\rho \bm{V} \right)  & = & 0 \label{eq:continuity_CDM}, \\
\bm{\nabla} \cdot \left( \bm{V} - f_\mathrm{p} \Delta \bm{V} \right)  & = & 0, \label{eq:incompress} \\
\frac{\mathrm{d} \bm{V}}{\mathrm{d}t}  & = & - 2  \bm{\Omega}_{\mathrm{K},0} \times \bm{V} + \left(\Omega^{2}_{\mathrm{K},0} - \Omega^{2}_\mathrm{K}\right)\bm{r} \nonumber \\
&& - \frac{\bm{\nabla} P }{\rho} + \bm{F}\left(\rho,f_\mathrm{p},\Delta \bm{V}\right) \label{eq:momentum_CDM}, \\
 \frac{\mathrm{d} \Delta \bm{V}}{\mathrm{d}t} & = &  - \frac{\Delta \bm{V}}{f_\mathrm{g} t_\mathrm{stop}} + \frac{\bm{\nabla P}}{f_\mathrm{g} \rho} - \left(\Delta \bm{V} \cdot \bm{\nabla} \right) \bm{V}  \nonumber  \\
 && +\bm{G}\left(f_\mathrm{p},\Delta \bm{V}\right) \label{eq:momentum_drift},
 \end{eqnarray}
 where
\begin{eqnarray} 
\frac{\mathrm{d}}{\mathrm{d}t}  & = & \frac{\partial }{\partial t} + \left( \bm{V} \cdot \bm{\nabla} \right), \\
\bm{F}\left(\rho,f_\mathrm{p},\Delta \bm{V}\right)  & = & - \frac{1}{\rho} \bm{\nabla} \cdot \left( f_\mathrm{p} \left(1-f_\mathrm{p}\right) \rho \Delta \bm{V} \otimes \Delta \bm{V} \right), \\
\bm{G}\left(f_\mathrm{p},\Delta \bm{V}\right)  & = &  f_\mathrm{p} \left(\Delta \bm{V}\cdot \bm{\nabla} \right) \left( f_\mathrm{p} \Delta \bm{V} \right) \nonumber \\
&& - f_\mathrm{g} \left(\Delta \bm{V}\cdot \bm{\nabla} \right) \left( f_\mathrm{g} \Delta \bm{V} \right).
\end{eqnarray}
The incompressibility condition \eqref{eq:incompress} reduces to $\mathrm{d} \rho = \mathrm{d} \rho_\mathrm{p}$, relating directly the dust over-concentration sought for to a local increase of the total density. This implies that gas can not accumulate locally. 

%-------------------------------------
\subsection{Local perturbations in a shearing box}

\subsubsection{Steady state solutions}
\label{sec:steady}

We use a cartesian shearing-box approximation $\left(\hat{\bm{x}}, \hat{\bm{y}}, \hat{\bm{z}} \right)$ \citep{Goldreich1965} and limit the study to local perturbations. Under this approximation, gas pressure can be decomposed into a background component, which consists of a small constant pressure force $g_{\rm e}$ and an additional perturbation. Denoting $H$ the pressure scale height of the gas, we have
\begin{equation}
g_\mathrm{e} \equiv -\frac{1}{\rho} \left. \frac{\partial P }{\partial r}\right|_{r_{0}} \sim \left( \frac{H}{r_{0}}\right)^{2} \Omega^{2}_{\mathrm{K},0} \, r_{0}  > 0  ,
\end{equation} 
since the disc is warmer and denser close to the star. The steady-state solution of Eqs.~\ref{eq:continuity_CDM} -- \ref{eq:momentum_drift} has been found by \citet{NSH1986}
\begin{eqnarray}
\bm{V}_0 & = & \left(-\frac{3}{2} \Omega_{\mathrm{K},0} x - \frac{g_\mathrm{e}}{2 \Omega_{\mathrm{K},0}} \right)\hat{\bm{y}}, \label{eq:v0_NSH}\\
\Delta \bm{V}_0  & = &  -\frac{g_\mathrm{e} t_\mathrm{stop}}{1 + S_\mathrm{t}^2} \hat{\bm{x}} + \frac{f_\mathrm{g} g_\mathrm{e} \Omega_{\mathrm{K},0} t_\mathrm{stop}^2}{2\left(1 + S_\mathrm{t}^2 \right)} \hat{\bm{y}} \label{eq:deltav0_NSH}.
\end{eqnarray}
To ease the forthcoming derivations, we adopt a definition of the Stokes number $S_\mathrm{t} \equiv  f_\mathrm{g} \Omega_K t_\mathrm{stop}$ that slightly differs from the usual notation by a factor $f_{\rm g}$. Eqs.~\ref{eq:v0_NSH} -- \ref{eq:deltav0_NSH} express that the motion is overall sub-Keplerian and that grains drift inwards towards high pressure regions, pushing gas outwards by angular momentum conservation. The drift velocity is the largest for Stokes numbers of order unity.

\subsubsection{Dimensionless quantities}

The natural timescale of the problem is the orbital timescale $\tau_0 \equiv \Omega^{-1}_{\mathrm{K},0} $. The physical length $\lambda_{\rm e}$ of the steady-state described in Sect.~\ref{sec:steady} is therefore $\lambda_\mathrm{e} \equiv g_\mathrm{e} /\Omega^{2}_{\mathrm{K},0} \sim (H/r_0) \, H \ll H$. $\lambda_{\rm e}$ gives the order of magnitude of the relative distance over which dust grains with $S_{\rm t} \sim 1$ and gas drift relatively to each other in a time $\tau_0$. Hence, we introduce the dimensionless time $\tau$, positions $\left( \chi, \zeta \right)$, and velocity $\bm{U}$ defined by
\begin{eqnarray}
t & \equiv &  \tau_0 \tau, \\
\left(x ,z \right) & \equiv &  \lambda_\mathrm{e} \! \left(\chi, \zeta \right), \\
\bm{V} & \equiv &  \frac{\lambda_\mathrm{e}}{\tau_0} \bm{U}, 
\end{eqnarray}
such as
\begin{eqnarray}
\bm{U}_0 & = &  - \left( \frac{3}{2} \chi + \frac{1}{2}\right) \hat{\bm{y}}, \label{eq:steady_adim1}\\
\Delta \bm{U}_0 & = &  - \left(\frac{S_\mathrm{t}}{f_\mathrm{g}\left(1+S_\mathrm{t}^2\right)}\hat{\bm{x}} - \frac{S_\mathrm{t}^2}{2 f_\mathrm{g} \left(1+S_\mathrm{t}^2\right)} \hat{\bm{y}} \right) \label{eq:steady_adim2}.
\end{eqnarray}

\subsubsection{Linear stability analysis}

We perform a linear perturbation analysis of Eqs.~\ref{eq:steady_adim1}
 -- \ref{eq:steady_adim2}, assuming a perturbation of the form

\begin{eqnarray}
\bm{U}  & = & \bm{U}_0 + \bm{u}(\tau,\chi,\zeta), \\
\Delta \bm{U}  & = & \Delta \bm{U}_0 + \Delta \bm{u}(\tau,\chi,\zeta), \\
\frac{\rho}{\rho_0}   & = & 1 + \delta(\tau,\chi,\zeta), \\
\frac{P - P_{0}}{\rho_0 g_\mathrm{e} \lambda_e}  & = & - \chi + h(\tau,\chi,\zeta), 
\end{eqnarray}
where $P_{0}$ denotes the pressure of the gas at the centre of the box. Following  \citet{Youdin2005,Jacquet2011}, the perturbation $f$ is decomposed under axisymmetric Fourier modes of the form 
\begin{equation}
f(\tau,\chi,\zeta) = \Tilde{f} \mathrm{e}^{\mathrm{i} \left(\kappa_x \chi + \kappa_z \zeta - \omega \tau \right)}.
\end{equation}
$\kappa_{x}$ should satisfy $\left| \kappa_{x} \right| \gg (H/r_0)^2$ to ensure consistency with the shearing-box approximation, and $\kappa_{z}$ should satisfy $\left| \kappa_{z} \right| \gg (H/r_0)$, to neglect the stratification of the disc. In practice, these conditions are not restrictive. The resulting set of equation in dimensionless form is
\begin{eqnarray}
-\mathrm{i} \omega \Tilde{\delta} + \mathrm{i} \bm{\kappa} \cdot \Tilde{\bm{u}} & = &  0, \label{eq:perturbedContinuity} \\
\bm{\kappa} \cdot \Tilde{\bm{u}} - f_\mathrm{p} \bm{\kappa} \cdot \Delta \Tilde{\bm{u}} + \kappa_x \frac{S_\mathrm{t}}{1 + S_\mathrm{t}^2} \Tilde{\delta} & = &  0,  \\
-\mathrm{i} \omega \Tilde{\bm{u}} - 2 \Tilde{u}_y \hat{\bm{x}} + \frac{1}{2}  \Tilde{u}_x \hat{\bm{y}}+ \Tilde{\delta} \hat{\bm{x}} + \mathrm{i} \bm{\kappa} \Tilde{h} +\Tilde{\bm{F}'} & = &  0, \\
-\mathrm{i} \omega \Delta \Tilde{\bm{u}} - 2 \Delta \Tilde{u}_y \hat{\bm{x}} + \frac{1}{2}  \Delta \Tilde{u}_x \hat{\bm{y}} - \mathrm{i} \frac{\bm{\kappa}}{f_\mathrm{g}} \Tilde{h}  +\nonumber \\
\frac{\Delta \Tilde{\bm{u}} + \Tilde{\delta} \Delta \bm{U}_0}{S_\mathrm{t}} - \mathrm{i} \kappa_x \frac{S_\mathrm{t}}{f_\mathrm{g}\left(1 + S_\mathrm{t}^2 \right)}\Tilde{\bm{u}} +\Tilde{\bm{G}'} & = &  0, \label{eq:perturbedDrift}
\end{eqnarray}
where
\begin{align}
\Tilde{\bm{F}'} & =   \mathrm{i} f_\mathrm{g} \left\lbrace \left( f_\mathrm{p} \bm{\kappa} \cdot \Delta \Tilde{\bm{u}} - \kappa_x \frac{S_\mathrm{t}}{1 + S_\mathrm{t}^2} \Tilde{\delta} \right) \Delta \bm{U}_0 -  \frac{f_\mathrm{p} \kappa_x S_\mathrm{t}}{f_\mathrm{g}\left(1 + S_\mathrm{t}^2\right)} \Delta \Tilde{\bm{u}}\right\rbrace,  \label{eq:perturbedF} \\
\Tilde{\bm{G}'} & =   - \mathrm{i} \kappa_x \frac{S_\mathrm{t}}{f_\mathrm{g}\left(1 + S_\mathrm{t}^2\right)} \left\lbrace \left(2 f_\mathrm{p} - 1 \right) \Delta \Tilde{\bm{u}} - f_\mathrm{g} \Delta \bm{U}_0 \Tilde{\delta} \right\rbrace. \label{eq:perturbedG}
\end{align}
Eqs.~\ref{eq:perturbedContinuity} -- \ref{eq:perturbedG} define a linear system of 8 equations on the 8 physical quantities $\Tilde{\delta}$, $\Tilde{\bm{u}}$, $\Tilde{h}$ and $\Delta \Tilde{\bm{u}}$ expressed in the above-defined dimensionless quantities \citep{Youdin2005}. A lengthy dispersion relation is obtained by setting to zero the polynomial determinant $\mathcal{P}_8$ of the system (see Appendix~\ref{sec:det}).
\vspace{-2 pt}
%________________________________________________________
\section{Unstable modes}
\label{sec:unstable}

\subsection{Reduced system: linear expansion in $S_{\rm t}$}
\label{sec:order1}

Since the expression of $\mathcal{P}_8$ is cumbersome, \citet{Youdin2005} and \citet{Jacquet2011} study alternatively a simplified set of equations by expanding Eqs.~\ref{eq:perturbedContinuity} -- \ref{eq:perturbedG} to the first order with respect to the Stokes number. The key idea brought by \citet{Youdin2005} and \citet{Jacquet2011} is to use the so-called terminal velocity approximation. Values at steady-state are used for the differential velocity between gas and dust for both the mean flow and the perturbation, assuming $S_{\rm t} \ll 1$ and performing the related Taylor expansion of the system. The resulting system is
\begin{eqnarray}
-\mathrm{i} \omega \Tilde{\delta} + \mathrm{i} \bm{\kappa} \cdot \Tilde{\bm{u}} = 0, \\
\mathrm{i} \bm{\kappa} \cdot \Tilde{\bm{u}} - \mathrm{i} \kappa_x S_\mathrm{t} \left(\frac{f_\mathrm{p}}{f_\mathrm{g}} - 1 \right) \Tilde{\delta} + f_\mathrm{p} \bm{\kappa}^2 \frac{S_\mathrm{t}}{f_\mathrm{g}} \Tilde{h} = 0, \\
-\mathrm{i} \omega \Tilde{\bm{u}} - 2 \Tilde{u}_y \hat{\bm{x}} + \frac{1}{2}  \Tilde{u}_x \hat{\bm{y}}+ \Tilde{\delta} \hat{\bm{x}} + \mathrm{i} \bm{\kappa} \Tilde{h} = 0, \\
\Delta \Tilde{\bm{u}} = \mathrm{i} S_\mathrm{t} \frac{\Tilde{h} \bm{\kappa}}{f_\mathrm{g}} + \frac{S_\mathrm{t}}{f_\mathrm{g}} \Tilde{\delta} \hat{\bm{x}} .\label{eq:perturbedterminalvelocity} 
\end{eqnarray}
One obtains the dispersion relation $P_\mathrm{Jac}(\omega) = 0$, with
\begin{align}
   P_\mathrm{Jac}(\omega)  \equiv  & \, S_\mathrm{t} \varepsilon \omega^4 + \mathrm{i} \omega^3 + S_\mathrm{t} \left( \mathrm{i} \kappa_{\rm x}  - \varepsilon \right) \omega^2 - \mathrm{i} \cos^2\theta \, \omega  \nonumber  \\ 
   &+ \mathrm{i}  \kappa_{\rm x} \cos^2\theta \left(\varepsilon-1 \right) S_\mathrm{t}, \label{eq:dispersionJacquet}
\end{align}
where $\cos \theta \equiv \kappa_z / \| \bm{\kappa} \|$ and $\varepsilon \equiv f_\mathrm{p}/f_\mathrm{g}$. Roots of Eq.~\ref{eq:dispersionJacquet} contains the secular mode $\omega_{\rm s}$ of the streaming instability
\begin{equation}
\omega_\mathrm{s} =  \frac{\kappa_x \left(f_\mathrm{p} - f_\mathrm{g} \right)}{f_\mathrm{g}} S_\mathrm{t} + \mathrm{o}(S_\mathrm{t} ) ,
\label{eq:oms}
\end{equation}
where the left-over $\mathrm{o}(S_\mathrm{t} ) $ of the right-hand side of Eq.~\ref{eq:oms} contributes at this order to the imaginary part as
\begin{equation}
\Im \left( \omega_\mathrm{s}  \right)  = \Im \left( \mathrm{o} \left( S_\mathrm{t} \right) \right) =   \left( \frac{\kappa_x^2}{f_\mathrm{g}^2} \frac{(f_\mathrm{p} - f_\mathrm{g})^2}{\cos^2 \theta}  \varepsilon \right) S_\mathrm{t}^3 = \mathcal{O}\left(S_{\rm t}^{3} \right) .
\label{eq:order3simp}
\end{equation}
Hence, \citet{Youdin2005} and \citet{Jacquet2011} infer a secular mode that is always unstable. The growth of the secular mode is interpreted by the mean of this reduced systems, by an interplay between drift towards pressure maxima, geostrophic balance and gas incompressibility. \citet{Youdin2005} also mention that epicycles are unstable as well when $\kappa_{z} \gg \kappa_{x}$. 

Similarly to \citet{Debras2020} -- Appendix~B -- we apply the argument theorem on the polynomial $\mathcal{P}_{\rm jac}$ to be more quantitative. We find that when $| \kappa_x | < S_\mathrm{t}  \kappa_z^2 \varepsilon$, $\mathcal{P}_{\rm jac}$ has 2 unstable roots, one corresponding to an approximated secular mode and the second one being a modified epicycle. This result \textit{on the reduced system} is exact (we verified it numerically). However, this criterion is \textit{incorrect} for describing the complete system of perturbed equations. Actually, numerical calculation of the roots of $\mathcal{P}_{8}$ shows that under the criterion derived above and for $\varepsilon < 1$, only the epicycle is unstable. Indeed, the reduced model is of order $S_{\rm t}$ and provides residuals of order $S_{\rm t}^{3}$. This strongly suggest that an expansion of order $S_{\rm t}^{3}$ is required to extract quantitatively the physics of the unstable modes of the streaming instability.

\subsection{Reduced system: third order expansion in $S_{\rm t}$} 
\label{sec:order3}

\subsubsection{Dispersion relation}

We perform an expansion of the system Eqs.~\ref{eq:perturbedContinuity} -- \ref{eq:perturbedG} to the order $S_{\rm t}^{3}$ and obtain an approximated dispersion relation $\mathcal{P}_{8}^{(3)}$. The detailed expression of $\mathcal{P}_{8}^{(3)}$ is lengthy and is given in App.~\ref{sec:app_order3}. The key idea is to  rearrange the terms via the Euclidian division that enforces a functional form that factorises the epicycles and the secular mode:
\begin{eqnarray}
    \frac{\mathcal{P}_{8}^{(3)} (\omega)}{\bm{\kappa}^2} & \equiv & \left(\omega - \left\lbrace \cos \theta + \alpha_1 S_\mathrm{t} + \alpha_2 S_\mathrm{t}^2  + \alpha_3 S_\mathrm{t}^3 \right\rbrace \right) \times \nonumber \\
    && \left(\omega - \left\lbrace- \cos \theta + \alpha_1 S_\mathrm{t} - \alpha_2 S_\mathrm{t}^2  + \alpha_3 S_\mathrm{t}^3 \right\rbrace \right) \times \nonumber \\
   && \left(\mathrm{i} \frac{S_\mathrm{t}^3}{f_\mathrm{g}} \omega^4 -\frac{2+f_\mathrm{g}}{f_\mathrm{g}} S_\mathrm{t}^2 \omega^3 + \beta_2 \omega^2 + \beta_1 \omega + \beta_0 \right) \nonumber \\
   &&+ S_\mathrm{t}^4\mathcal{R}^{(3)}(\omega),  
\label{eq:p3}
\end{eqnarray}    
The residual $\mathcal{R}^{(3)}$ is a polynomial of degree $5$ such that $S_\mathrm{t}^4\mathcal{R}^{(3)}(\omega)$ is of order $S_\mathrm{t}^4$ when $\omega \lesssim 1$, and has therefore negligible contribution \textit{per} construction. The coefficients $\alpha_{1,2,3}$ and $\beta_{0,1,2}$ are given in Appendix~\ref{sec:coeffs_p3}. The conditions of validity for the aforesaid expansion are $\cos^2 \theta \gg S_{\rm t}^2$, $\kappa_{x}^2 S_{\rm t} \lesssim 1$ and $\kappa_{z} S_{\rm t} \lesssim 1 $.

We note that performing the same technic while restraining the expansion to the first order in $S_{\rm t}$ gives an approximate dispersion relation under the form
\begin{equation}
    \mathrm{i} \frac{\mathcal{P}_{8}^{(1)} (\omega) }{\bm{\kappa^2}} \equiv \left( 1-3\mathrm{i} \omega S_\mathrm{t} \right) P_\mathrm{Jac}(\omega) + S_\mathrm{t}^2 \mathcal{R}^{(1)}(\omega) = 0,
    \label{eq:link}
\end{equation}
with
\begin{equation}
    \mathcal{R}^{(1)}(\omega) \equiv 3 \left\lbrace \mathrm{i} \varepsilon \omega^5 - \left(\mathrm{i} \varepsilon + \kappa_x \right) \omega^3 + \cos^2 \theta \left(1- \varepsilon \right)\omega \right\rbrace .
\end{equation}
Eq.~\ref{eq:link} demonstrates that the expansion of Sect.~\ref{sec:order1} is actually a first order expansion in $S_{\rm t}$, although it was not mentioned explicitly in previous studies. Eq.~\ref{eq:link} shows that for a set of parameters that maximises the growth rate,  $\omega_{\rm s}$ is of order $S_{\rm t}$ and the model presented in Sect.~\ref{sec:order1} is accurate. This is actually the choice of parameters chosen by \citet{Youdin2005}, certainly adopted to highlight the efficiency of the instability. This choice of parameter may explain why stability of the secular mode has been over-looked so far. When $\omega_{\rm s}$ is \textit{not} of order $S_{\rm t}$, Eq.~\ref{eq:link} shows that a linear expansion fails to describe quantitatively the evolution of the perturbations. 

\subsubsection{Secular mode}
\label{sec:secular3}

Eq.~\ref{eq:p3} provides directly the expression $\omega_\mathrm{s}$ of the frequency of the secular mode at third order with respect to $S_{\rm t}$ as
\begin{equation}
    \omega_\mathrm{s} = \frac{\kappa_x \left(f_\mathrm{p} - f_\mathrm{g} \right)}{f_\mathrm{g}} S_\mathrm{t} + \omega_\mathrm{s}^{(3)} S_\mathrm{t}^3 + \mathcal{O}\left(S_{\rm t}^{4} \right),
\end{equation}
with 
\begin{equation}
    \Im ( \omega_\mathrm{s}^{(3)} ) = \frac{\kappa_x^2}{f_\mathrm{g}^2}\left( \frac{(f_\mathrm{p} - f_\mathrm{g})^2}{\cos^2 \theta}\varepsilon + 3 f_\mathrm{p}\left(f_\mathrm{p} - f_\mathrm{g} \right) \right).
 \label{eq:oms3_full}
\end{equation}
The imaginary part of $ \omega_\mathrm{s}^{(3)}$ is now consistently expressed up to the order $S_{\rm t}^{3}$. This correction differs from the one obtained by a linear expansion Eq.~\ref{eq:order3simp} by its last term. The extra contribution originates from the terms $\left(\Delta \bm{U} \cdot \nabla \right) \bm{u}$ that corresponds to the differential advection of the perturbations by the gas and the dust. In a linear approximation, the contribution of the back-reaction to the mean flow is negligible at order $S_{\rm t}$ \citep{NSH1986}. However, this correction becomes important at order $S_{\rm t}^{3}$. Eq.~\ref{eq:oms3_full} shows effects of back-reaction onto the drift are significant in the regime where the streaming instability develops and must be accounted for. When $\kappa_{x} \gg \kappa_{z}$, Eq.~\ref{eq:oms3_full} reduces to the analysis of \citet{Youdin2005,Jacquet2011}.

Eq.~\ref{eq:oms3_full} shows that this correction is critical to understand the development of the secular mode of the streaming instability. Indeed, the secular mode can be \textit{stable} when the conditions
\begin{eqnarray}
    f_{\rm p} & < & f_{\rm g} , \\
    | \kappa_x | & \leq &  \sqrt{2} | \kappa_z | ,
\end{eqnarray}
are satisfied. If not, the secular mode becomes unstable when
\begin{equation}
\left| \kappa_x \right| > | \kappa_z | \sqrt{\frac{2 f_{\rm g} + f_{\rm p}}{f_{\rm g} - f_{\rm p}}}   \ge \sqrt{2} | \kappa_z | ,
    \label{eq:critsec}
\end{equation}
with equality when $f_{\rm p} = 0$. If $f_{\rm p} >f_{\rm g}$, the secular mode is always unstable as evidenced by \citet{Youdin2005}.

\begin{figure}
	\centering{\includegraphics[width=\columnwidth]{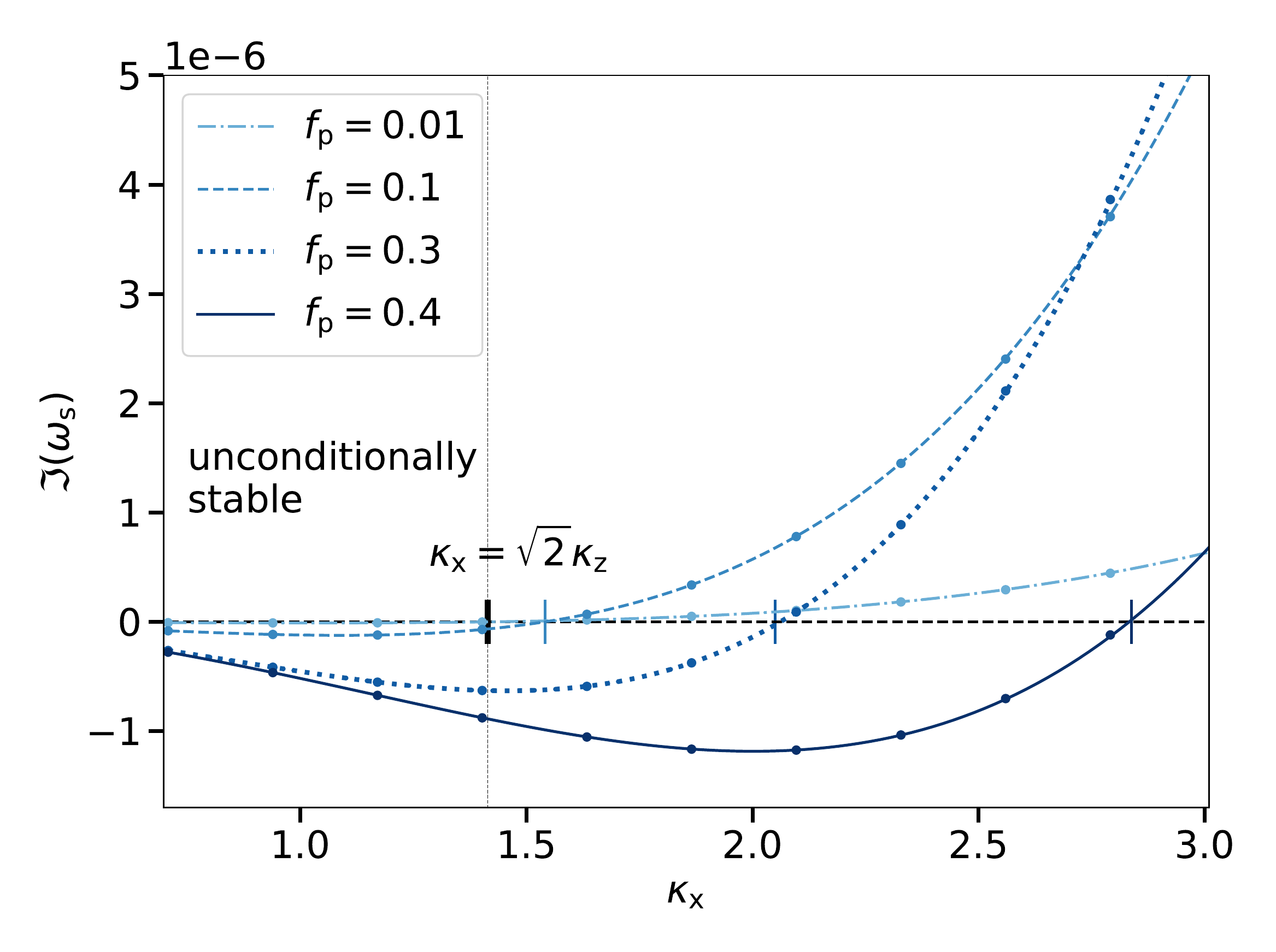}}
	\caption{Imaginary part of the secular mode of the streaming instability calculated numerically from the full set of hydrodynamical equations $\mathcal{P}_{8}$ for dust fractions of $f_{\rm p} = 0.01, 0.1, 0.3, 0.4$ (from light to dark blue lines respectively). The secular mode of the streaming instability is \textit{always stable} when $f_{\rm p} < f_{\rm g}$ and  $| \kappa_x | \leq \sqrt{2} | \kappa_z |$. Dots indicate the corresponding values predicted by the third-order expansion $\mathcal{P}_{8}^{(3)}$. The agreement is almost perfect. Here, the Stokes number is fixed to $S_{\rm t} = 0.01$ and $\kappa_{z} = 1$.}
	\label{fig:stable_sec}
\end{figure}

Fig.~\ref{fig:stable_sec} illustrates this property by showing the imaginary part of $\omega_{s}$ obtained from a direct numerical resolution of the roots of the full dispersion relation $\mathcal{P}_{8}$. The roots obtained by the third-order expansion $\mathcal{P}_{8}^{(3)}$ are displayed as well, both of them showing almost perfect agreement. We fix $S_{\rm t} = 0.01$ to show that substantial corrections to the linear model can be obtained even for small grains. We then set $\kappa_{z} = 1$ and vary the dust fraction according from $f_{\rm p} = 0.01$ to $f_{\rm p} = 0.4$. When the criterion $ | \kappa_x | \leq \sqrt{2} | \kappa_z |$ is satisfied, the imaginary part of $\omega_{\rm s}$ is always negative and the secular mode is stable, as expected. For  $| \kappa_x | > \sqrt{2} | \kappa_z |$, it becomes unstable when the condition of Eq.~\ref{eq:critsec} is satisfied. The related critical values of $| \kappa_x |$ increase with $f_{\rm p}$, as predicted by Eq.~\ref{eq:critsec}. 

\subsubsection{Unstable epicycles}

Eq.~\ref{eq:p3} shows that epicycles can be unstable as well since
\begin{align}
\Im \left( \omega_{\rm e} \right) & = \frac{\varepsilon}{2} S_{\rm t} \left\lbrace   \frac{ | \kappa_x | \varepsilon }{\cos \theta} S_{\rm t} -  \kappa_x^2 \left( 1  + 3 \varepsilon + \frac{(\varepsilon -1)^2}{\cos^2 \theta}\right) S_{\rm t}^2  \, -   \sin^2 \theta \, \times \right. \nonumber \\
& \left.  \left( 1 + \frac{ | \kappa_x | (\varepsilon + 2) }{2 \cos \theta} S_{\rm t} + \left( \varepsilon (2 \varepsilon + 3 )\sin^2 \theta - (\varepsilon +1)^2  \right) S_{\rm t}^2 \right) \right\rbrace .
\label{eq:im_epi}
\end{align}
\begin{figure}
	\centering{\includegraphics[width=\columnwidth]{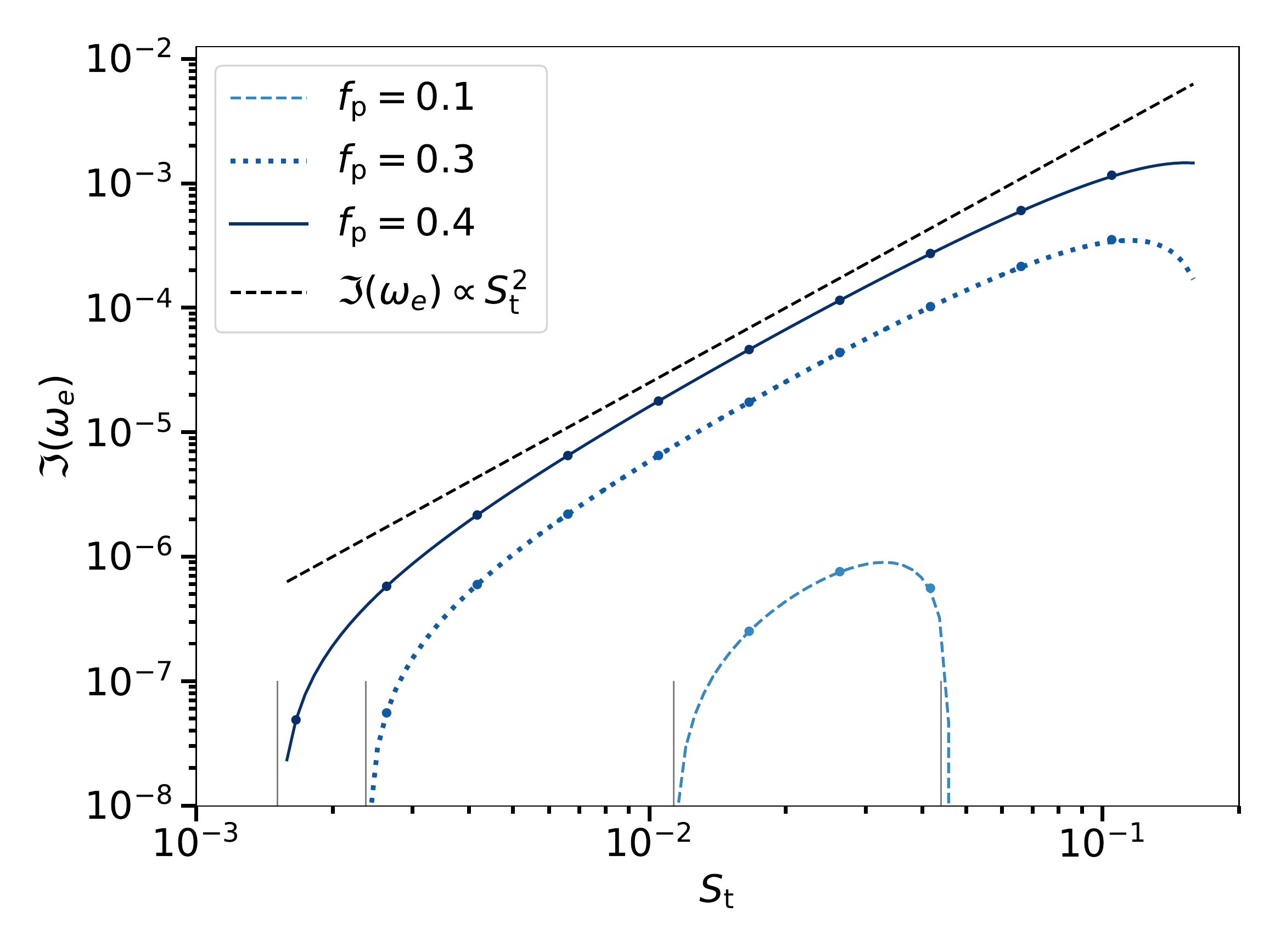}}
	\caption{Imaginary part of the unstable epicycle of the streaming instability calculated numerically from the full set of hydrodynamical equations $\mathcal{P}_{8}$ for dust fractions of $f_{\rm p} = 0.1, 0.3, 0.4$ (from light to dark blue lines respectively) for varying Stokes numbers. The growth rates varies as $S^{2}_{\rm t}$, as expected.  Dots indicate the corresponding values predicted by the third-order expansion $\mathcal{P}_{8}^{(3)}$ and shows almost perfect agreement. The vertical solid gray lines indicate the analytic stability criterion given by Eq.~\ref{eq:epi_unstable}. We choose $\kappa_{x} = 0.9$ and $\kappa_{z} = 30$ for the secular mode to be  stable.}
	\label{fig:epi}
\end{figure}
 The epicycle becomes therefore unstable under the necessary but unrestrictive condition $\sin^2 \theta \ll 1$. A reasonable approximation for instability is derived by expanding Eq.~\ref{eq:im_epi} to the third order in $\sin \theta$. One obtains
\begin{eqnarray}
| \kappa_x | \varepsilon S_{\rm t} &\geq& \sin^2 \theta + \kappa_x^2 \left( 1 +  (\varepsilon +1)^2 \right) S_{\rm t}^2 .
\label{eq:epi_unstable}
\end{eqnarray}
To first order in Stokes, this criterion reduces to $| \kappa_x | \leq  S_{\rm t} \kappa_z^2 \varepsilon$, as found in Sect.~\ref{sec:order1}. Eq.~\ref{eq:im_epi} shows that the unstable epicycle growth scales as $S_{\rm t}^{2}$. Fig.~\ref{fig:epi} shows the imaginary part of the unstable epicycle obtained numerically from the complete dispersion relation $\mathcal{P}_{8}$. The agreement with the analytic expansion is almost perfect. In particular, the analytic stability criterion given by Eq.~\ref{eq:epi_unstable} is well satisfied. Remarkably, the growth of the epicycle can occur in a few $10^{3}$ of orbital periods for $f_{\rm p} \gtrsim 0.2$ and $S_{\rm  t} \gtrsim 0.01$, a relevant timescale for planetesimal formation (see Fig.~\ref{fig:omegaes}).

\subsubsection{Epicycles vs. secular mode}

\begin{figure}
	\centering{\includegraphics[width=\columnwidth]{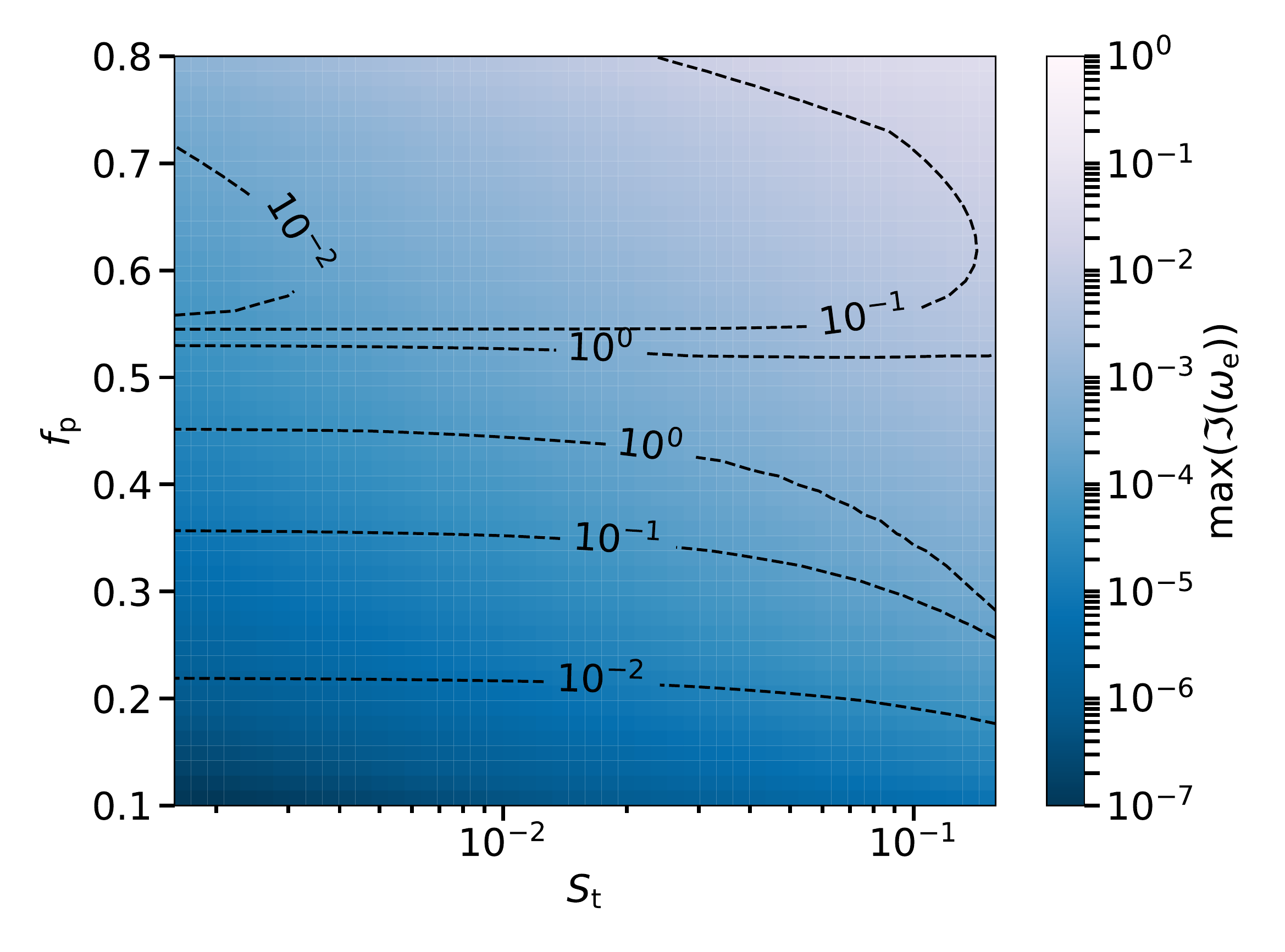}}
	\caption{Color map: maximum growth rate of the unstable epicycle, varying $\kappa_{x}$ and $\kappa_{z}$ within the range $\left[ 0.1 ; S_{\rm t}^{-1} \right]$. Typical growth times of $ \sim 10^{3}$ orbital periods are obtained for $f_{\rm p} \gtrsim 0.2$ and $S_{\rm  t} \gtrsim 0.01$. Dashed black contours: ratio between the maximum growth rates of the epicycle vs. the secular mode. These do not correspond to the same $\kappa_{x} $ and $\kappa_{z} $ \textit{a priori}. The epicycle mode can grow as fast as the secular mode.}
	\label{fig:omegaes}
\end{figure}

An indicator of the relative efficiency of the two unstable modes can be obtained by the following procedure. For each mode and a given value of $f_{\rm p}$ and $S_{\rm t}$, one maximises the growth rate with respect to $\kappa_{x}$ and $\kappa_{z}$. The ratio of the values obtained for the two modes are then compared, keeping in mind that maxima are not reach for the same values of $\kappa_{x}$ and $\kappa_{z}$ \textit{a priori}. For consistency with the shearing-box approximation and the expansion of Sect.~\ref{sec:order3}, $\kappa_{x}$ and $\kappa_{z}$ are chosen in the range $\left[ 0.1 ; S_{\rm t}^{-1} \right]$. Fig.~\ref{fig:omegaes} shows that the growth rate of the epicycle can be as large as the one of the secular mode, for a wide range of dust fractions and Stokes numbers relevant for planetesimal formation. 

To interpret this result, one first finds approximations for the values of $\kappa_{x}$ and $\kappa_{z}$ that maximises the growth rate of the epicycle:
\begin{eqnarray}
| \kappa_z | &\simeq& S_{\rm t}^{-1} ,\\
| \kappa_x | &\simeq& \frac{\varepsilon}{2\left( 2 + (\varepsilon+1)^2 \right)} | \kappa_z |, \, \text{if } f_{\rm p}<f_{\rm g}, \nonumber \\
| \kappa_x | &\simeq& \frac{\varepsilon}{2\left( 1+ (\varepsilon+1)^2 \right)} | \kappa_z |, \, \text{if } f_{\rm p} \geq f_{\rm g}.
\end{eqnarray}
The above dependancy in $S_{\rm t}^{-1}$ for $\kappa_{x}$ was originally commented by \citet{Youdin2005} -- their Short-Wavelength limit -- but without mathematical justification. Similarly, for the secular mode, one obtains $\kappa_x \sim S_{\rm t}^{-1/2}$ as \citet{Youdin2005}. In particular, one can explain the ridge observed in Fig.~\ref{fig:omegaes} for the contour line corresponding to $10^{0}$, i.e. similar growth rate for the two modes. On one hand, the secular mode becomes stable for $f_{\rm p} = f_{\rm g} = 0.5$. On the other hand, the secular mode approaches the marginal limit of equality in Eq.~\ref{eq:critsec} for $S_{\rm t} \gtrsim 3 \, 10^{-2}$ and $f_{\rm p}  \lesssim 0.5$. Indeed, $\kappa_{x} \sim S_{\rm t}^{-1/2}$ and $\kappa_{z}$ is bounded by the value $0.1$. Importantly, the corrections of order 3 introduced in Sect.~\ref{sec:order3} are necessary to interpret the appearance of this ridge. Finally, the phase velocity of the fastest growing epicycle matches the radial drift velocity of the background in the limit $f_{\rm p} \ll 1$ as found by \citet{Squire2018}.

\section{Relevance for planetesimal formation}
\label{sec:discuss}

\subsection{Viscous damping}
\label{sec:viscous}

As a rule of thumb, one can estimate the resilience of the unstable modes with respect to viscous damping by comparing the viscous timescale and the typical time over which the instability develops. In dimensionless quantities, this condition yields $\Im(\omega) \tau_\nu  \gtrsim 1$, where 
\begin{align}
\tau_\nu = \frac{4 \pi^2}{\bm{\kappa}^2} \frac{1}{\alpha} \frac{g_{\rm e}^2}{\Omega_{\mathrm{K},0}^2 c_{\rm s}^2} \sim \frac{4 \pi^2}{\bm{\kappa}^2} \frac{1}{\alpha} \left( \frac{H}{r_0}\right) .
\end{align}
Instability resists viscosity when $ \Im(\omega) \gtrsim \alpha \frac{\bm{\kappa}^2}{4 \pi^2}  \left( \frac{r_0}{H}\right)$. For typical discs with $\alpha = 5 \times 10^{-4}$ and $H/r_0 = 0.1$,  one gets $ \Im(\omega) \gtrsim  10^{-4} \bm{\kappa}^2$. For the secular mode, the validity of this condition has been discussed in several studies (e.g. \citealt{Youdin2005,Auffinger2018}). For $\alpha \gtrsim 10^{-5} - 10^{-4}$, the growth of the secular mode is damped.

More generally, the secular mode grows when $ \kappa_{x} \gtrsim \kappa_{z}$, implying that he threshold for viscous damping is set by the value of $\kappa_{x}$. Moreover, the growth rate of the secular mode varies as $\sim \varepsilon \kappa_{x}^{2} S_{\rm t}^{3} / \cos^{2}\theta$ (Eq.~\ref{eq:oms3_full}, see also \citealt{Youdin2005,Jacquet2011}). Large growth rates could be achieved with large values of $\kappa_{x}$. However, those modes are damped by viscosity. One finds that no secular mode can develop for $S_{\rm t} \lesssim 0.01$. For values of $S_{\rm t}$ increasing from $0.01$ to $0.1$, only secular modes with reasonably small values of $\kappa_{x}$ can develop and the associated timescales go from $\sim 10^{6}$ to $\sim 10^{3}$ orbital periods. This regime becomes therefore relevant for planetesimal formation for $S_{\rm t} \gtrsim 0.1$. 

\begin{figure}
	\centering{\includegraphics[width=\columnwidth]{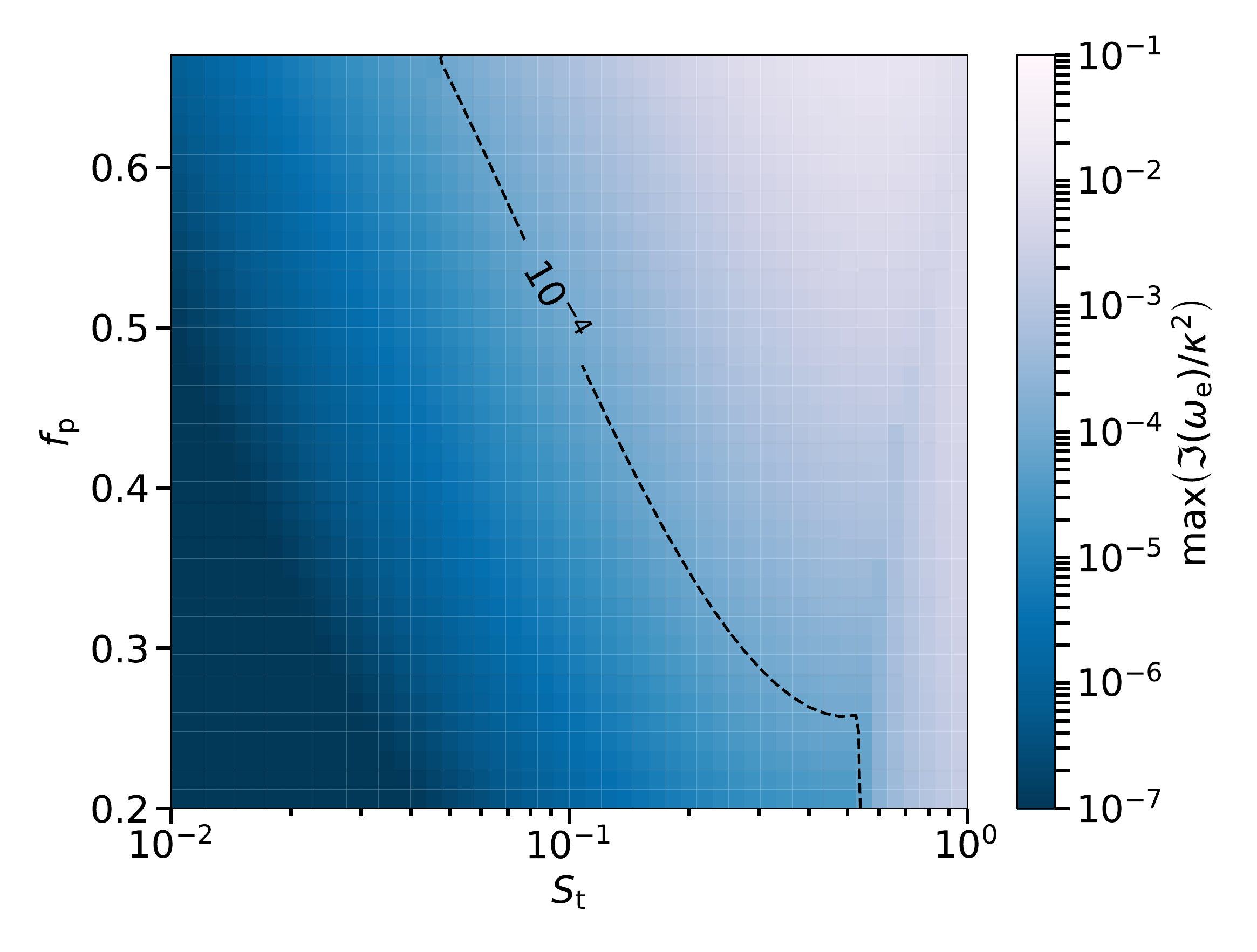}}
	\caption{Color map: maximum growth rate of the unstable epicycle normalised to $\kappa^{2}$. For a value $\alpha$ larger than 5 times the value indicated by the blue color bar, the instability is damped by viscous dissipation. As an example, above the dashed black contour labeled $10^{-4}$, the epicycle is unstable and resists the viscous damping associated to a value of $\alpha = 5 \times 10^{-4}$. Hence, epicycles can be more resilient against viscosity compared to the secular mode. Here, the aspect ratio is $H/r_{0} = 0.1$.}
	\label{fig:omegaevisc}
\end{figure}

Fig.~\ref{fig:omegaevisc} shows similar analysis for the unstable epicycle in a disc where $\alpha = 5 \times 10^{-4}$ and $H/r_0 = 0.1$. From Sect.~\ref{sec:order3}, one knows that epicycles  become unstable for $\kappa_{z} \gg \kappa_{x}$. For this mode, the threshold for viscous damping is thus set by the value of $\kappa_{z}$. On the other hand, Eq.~\ref{eq:im_epi} shows that the growth rate of the epicycle varies as $\sim \epsilon^{2} \kappa_{x} S_{\rm t}^{2} $. Reasonably small values of $\kappa_{z}$ and $\kappa_{x}$ can therefore allow the instability to develop without being damped. This happens for $S_{\rm t} \gtrsim 0.1$ and gives timescales of $\sim 10^{3}$ orbital periods, which compares with the ones obtained for the secular mode. For both modes, much shorter growth time can be achieved for larger Stokes numbers (see Fig.~\ref{fig:omegaes}). For classical T-Tauri star discs, streaming instability may therefore concentrate efficiently (sub-)millimetre-in-size grains, relieving the constrain of the fragmentation barrier. As a final remark, \citet{Auffinger2018} have shown that around a pressure bump, streaming instability may favours epicycles with respect to the secular mode (in this case, for large amplitudes of the bump) and resists viscous damping. In this situation as well, epicycles can not be neglected. 

\subsection{Green's function analysis}

In real discs, power spectrum is expected to peak at the orbital frequency and to cascade down by turbulence to larger frequencies. Hence, power is essentially injected at frequencies close to one of the epicycles. To understand how a dusty discs responds to a local perturbation, we study the evolution of a perturbation (Eqs.~\ref{eq:perturbedContinuity} -- \ref{eq:perturbedG}) to a monochromatic source, switched on at $\tau = 0$, of the form 
\begin{eqnarray}
\bm{S_8}(\tau, \chi, \zeta) =  \Theta(\tau)  \Tilde{\bm{S_8}} \mathrm{e}^{\mathrm{i} \left(\kappa_x \chi + \kappa_z \zeta - \omega_{\rm f} \tau \right)} .
\label{eq:source}
\end{eqnarray}
$\omega_{\rm f}$ denotes the real driving frequency of the source, $\Theta(\tau)$ the Heaviside step function, $\bm{S_8}(\tau, \chi, \zeta)$ the vector expression of the source and $\Tilde{\bm{S_8}}$ its Fourier decomposition. Both have eight components corresponding to the eight perturbed quantities ($\delta$, $\bm{u}$, $\Delta \bm{u}$, $h$).
With these notations, the system of perturbed equations writes
\begin{eqnarray}
\left( \bm{\Pi}_7 \partial_\tau +  \textbf{M}_8(\kappa_x, \kappa_z) \right) \bm{P} = \bm{S_8}(\tau, \chi, \zeta),
\end{eqnarray}
where $\bm{\Pi}_7 = \text{Diag}(1,1,1,1,1,1,1,0)$  and the matrix of perturbations $\textbf{M}_8(\kappa_x, \kappa_z)$ are 8x8 matrices, and $\bm{P} =  (\delta, \bm{u}, \Delta \bm{u}, h)$ is a vector with eight components.
Using a Laplace-transform and applying the residue theorem (e.g. \citealt{Morse1953}), one obtains
\begin{align}
\Tilde{\bm{P}}(\tau) & = \Theta(\tau)  \left(  \left[ - \mathrm{i}\omega_{\rm f}  \bm{\Pi}_7  +  \textbf{M}_8(\kappa_x, \kappa_z)  \right]^{-1} \Tilde{\bm{S_{8}}}  \mathrm{e}^{ - \mathrm{i}\omega_{\rm f} \tau } \right. \nonumber \\
& \left. + \sum_{n=1}^6  \frac{ \mathrm{adj} \left\lbrace -\mathrm{i}\omega_{n}  \bm{\Pi}_7  +  \textbf{M}_8(\kappa_x, \kappa_z)  \right\rbrace \Tilde{\bm{S_{8}}} }{- \mathrm{i} \partial_{\omega} \mathcal{P}_8 (\omega_{\mathbb{R},n} + \mathrm{i}s_n)}  \frac{ \mathrm{e}^{\left(s_n \tau - \mathrm{i}\omega_{\mathbb{R},n} \tau \right)}}{\mathrm{i}(\omega_{\rm f} -\omega_{\mathbb{R},n}) + s_n }    \right)  ,
\label{eq:linear_response}
\end{align}
where $\bm{P}(\tau,\chi,\zeta) = \Tilde{\bm{P}}(\tau)\mathrm{e}^{\mathrm{i} \left(\kappa_x \chi + \kappa_z \zeta \right)}$ is the response of the disc to the excitation $\bm{S_8}(\tau, \chi, \zeta)$, $\omega_n = \omega_{\mathbb{R}, n} + \mathrm{i} s_n$ is the $n$-th zero of the dispersion relation $\mathcal{P}_8(\omega)=0$ and $\mathrm{adj} $ denotes the matrix adjugate. The form of Eq.~\ref{eq:linear_response} is generic. Similar responses have been extensively studied in the literature (e.g. \citealt{Huerre1990,Lingwood1997}).

From Eq.~\ref{eq:linear_response}, the perturbation can be decomposed in two parts: an oscillatory part with frequency $\omega_{\rm f}$ (the first term of the right-hand side of Eq.~\ref{eq:linear_response}) and a superposition of the six characteristic waves of the disc that may grow or be damped. Would all waves be damped, the asymptotic response at large times would reduce to the single usual oscillatory part of frequency $\omega_{\rm f}$. The interesting part for planetesimal formation is the transient regime described by the second term of the right-hand side of Eq.~\ref{eq:linear_response}, which is dominated by growing modes. Streaming instability requires care, since two unstable modes with similar growth rates coexist (Sect.~\ref{sec:viscous}). 

The source term $\bm{S_8}$ excites the waves with different amplitudes. Eq.~\ref{eq:linear_response} shows that these amplitudes result from cumulative effects due to different factors.  A first factor of spatial origin is the decomposition of the source term onto the eigen-vectors of $ \mathrm{adj} \left\lbrace -\mathrm{i}\omega_{n}  \bm{\Pi}_7  +  \textbf{M}_8(\kappa_x, \kappa_z)  \right\rbrace$. In real discs, source terms are stochastic and should not favour any eigen-mode in average. We expect therefore a similar mean contribution of this factor for both the epicycles and the secular mode. A second factor of temporal origin is the product $ (\mathrm{i}(\omega_{\rm f} -\omega_{\mathbb{R},n}) + s_n) \partial_{\omega} \mathcal{P}_8 (\omega_{\mathbb{R},n} + \mathrm{i}s_n)$, which combines the distance of the driving frequency to the frequency of the unstable modes, and the ability of the disc to respond at the waves frequencies. Importantly, power is preferentially injected at frequencies $\omega_{\rm f}$ close to the epicyclic frequencies. We therefore study the response to excitations such as $\omega_{\rm f} \sim \omega_{\mathbb{R},\rm e} \sim 1$.

We use the analytic expression for $\mathcal{P}_{8}^{(3)}$ derived in Sect.~\ref{sec:order3} to estimate the relative values of the factors $\partial_{\omega} \mathcal{P}_8 (\omega_{\rm e, s})$ that weight the driven amplitudes $a_{\rm e}$ and $a_{\rm s}$ of the epicycles and the secular modes respectively. The driving term $\bm{S_8}$ has been decomposed onto spatial Fourier mode in Eq.~\ref{eq:source} and the driving frequency $\omega_{\rm f}$ can be associated to several values of $\bm{\kappa}$, themselves associated to various epicycles and secular modes. For both modes, we obtain the relation $\partial_{\omega} \mathcal{P}_8 (\omega_{\rm e, s}) \sim \bm{\kappa}^2 \left(\cos^2 \theta + \mathcal{O}\left(S_{\rm t} \right) \right) $. Hence the scalings $\partial_{\omega} \mathcal{P}_8 (\omega_{\rm e}) \sim \bm{\kappa}^2$ for unstable epicycles and  $\partial_{\omega} \mathcal{P}_8 (\omega_{\rm s}) \sim \bm{\kappa}^2 S_{\rm t}$ for the secular mode. Unstable epicycles should additionally satisfy $s_{\rm e} \sim S_{\rm t}^2$ and $\sin \theta \lesssim S_{\rm t}$ (Sect.~\ref{sec:order3}). One gets
\begin{equation}
\mathrm{i}(\omega_{\rm f} -\omega_{\mathbb{R},\rm e}) + s_{\rm e} \simeq s_{\rm e} \sim S_{\rm t}^2 .
\end{equation}
On the other hand, the secular mode satisfies
\begin{equation}
\mathrm{i}(\omega_{\rm f} -\omega_{\mathbb{R},\rm s}) + s_{\rm s} \simeq \mathrm{i} \, \omega_{\rm f} \sim 1 .
\end{equation}
Combining all these contributions gives a ratio
\begin{equation}
\frac{a_{\rm e}}{a_{\rm s}} \sim \frac{S_{\rm t}^{-2}}{S_{\rm t}^{-1}}\frac{\bm{\kappa}_{\rm s}^2} {\bm{\kappa}_{\rm e}^2} ,
\label{eq:ampl_ratio_in}
\end{equation}
for the relative amplitudes of the epicycles and the secular mode. From Sect.~\ref{sec:viscous}, the values of $\bm{\kappa}_{\rm s}^2$ and $\bm{\kappa}_{\rm e}^2$ that ensures for the modes to resist viscous damping and to develop in timescales relevant for planetesimal formation are such that $\bm{\kappa}_{\rm s}^2 / \bm{\kappa}_{\rm e}^2 \sim 1$. Hence, Eq.~\ref{eq:ampl_ratio_in} reduces to     
\begin{equation}
\frac{a_{\rm e}}{a_{\rm s}}  \sim S_{\rm t}^{-1}  \gg 1,
\label{eq:ampl_ratio}
\end{equation}
We therefore expect that for pebbles with $S_{\rm t} \ll 1$, streaming instability develops in discs essentially though the channel of its unstable epicycles. At later times, the secular mode will assist the growth, but non-linear effects may already be not negligible anymore.
\section{Conclusion}
\label{sec:conclu}

In this study, we revisit the linear growth of the streaming instability in dusty discs. The dispersion relation that characterises linear perturbations is analysed by the mean of a self-consistent expansion at third order with respect to the Stokes number. We provide an approximate dispersion relation that factorises the two epicycles and the secular mode. Important terms that were neglected previously are subsequently integrated. The analytic approximation agrees almost perfectly with numerical results on the full system. Moreover, we use Green's function analysis to investigate the response of a disc to realistic excitations. From these derivations, we find that:
\begin{enumerate}
\item Contrary to what is often mentioned in the literature, the secular mode can be stable. We derive an accurate analytic criterion for its stability (Eq.~\ref{eq:critsec}).
\item Epicycles can also be unstable, whether the secular mode is stable or not. We derive its growth rates and its associated stability condition (Eq.~\ref{eq:epi_unstable}).  
\item Epicyclic modes can grow as fast as the secular modes. They can however be more resilient against viscous damping and be excited most efficiently in real discs (Eq.~\ref{eq:ampl_ratio}).
\end{enumerate}
Streaming instability is known to be a privileged mechanism for planetesimal formation, but from the findings of this study, it may preferentially develop through the unexpected channel of unstable epicycles.

\section*{Acknowledgements}
We acknowledge financial support from the national programs (PNP, PNPS, PCMI) of CNRS/INSU, CEA, and CNES, France. This project was partly supported by the IDEXLyon project (contract nANR-16-IDEX-0005) under the auspices University of Lyon. This project has received funding from the European Union's Horizon 2020 research and innovation programme under the Marie Sk\l odowska-Curie grant agreement No 823823. This work has received funding from the European Research Council (ERC) under the European Union’s Horizon 2020 research and innovation programme (grant agreement ERC advanced grant 740021–ARTHUS, PI: Thomas Buchert). We thank the referee for his constructive report. We have use \textsc{Mathematica} \citep{Mathematica}

\bibliographystyle{mnras}
\bibliography{Streaming}

%%%%%%%%%%%%%%%%%%%%%%%%%%%%%%%
%%%%%%%%%% APPENDICES %%%%%%%%%
%%%%%%%%%%%%%%%%%%%%%%%%%%%%%%%

\newpage
\appendix

\section{Determinant of the linear system $\mathcal{P}_{8}$}
\label{sec:det}

\begin{multicols}{1}
\begin{equation}
\begin{small}
\begin{vmatrix}
 -\mathrm{i} \omega  & \mathrm{i} \kappa_x & 0 & \mathrm{i} \kappa_z & 0 & 0 & 0 & 0 \\
 \frac{\mathrm{i} \kappa_x S_\mathrm{t}^2}{\left(S_\mathrm{t}^2+1\right)^2}+1 & -\mathrm{i} \omega  & -2 & 0 & -\frac{2 \mathrm{i} f_\mathrm{p} \kappa_x S_\mathrm{t}}{S_\mathrm{t}^2+1} & 0 & -\frac{\mathrm{i} f_\mathrm{p} \kappa_z S_\mathrm{t}}{S_\mathrm{t}^2+1} & \mathrm{i} \kappa_x \\
 -\frac{\mathrm{i} \kappa_x S_\mathrm{t}^3}{2 \left(S_\mathrm{t}^2+1\right)^2} & \frac{1}{2} & -\mathrm{i} \omega  & 0 & \frac{\mathrm{i} f_\mathrm{p} \kappa_x S_\mathrm{t}^2}{2 \left(S_\mathrm{t}^2+1\right)} & -\frac{\mathrm{i} f_\mathrm{p} \kappa_x S_\mathrm{t}}{S_\mathrm{t}^2+1} & \frac{\mathrm{i} f_\mathrm{p} \kappa_z S_\mathrm{t}^2}{2 \left(S_\mathrm{t}^2+1\right)} & 0 \\
 0 & 0 & 0 & -\mathrm{i} \omega  & 0 & 0 & -\frac{\mathrm{i} f_\mathrm{p} \kappa_x S_\mathrm{t}}{S_\mathrm{t}^2+1} & \mathrm{i} \kappa_z \\
 \frac{\mathrm{i} \kappa_x S_\mathrm{t}^3}{f_\mathrm{g} \left(S_\mathrm{t}^2+1\right)^2}-\frac{S_\mathrm{t}}{f_\mathrm{g} \left(S_\mathrm{t}^2+1\right)} & -\frac{\mathrm{i} \kappa_x S_\mathrm{t}^2}{f_\mathrm{g} \left(S_\mathrm{t}^2+1\right)} & 0 & 0 & -\frac{\mathrm{i} (2 f_\mathrm{p}-1) \kappa_x S_\mathrm{t}^2}{f_\mathrm{g} \left(S_\mathrm{t}^2+1\right)}-\mathrm{i} \omega  S_\mathrm{t}+1 & -2 S_\mathrm{t} & 0 & -\frac{\mathrm{i} \kappa_x S_\mathrm{t}}{f_\mathrm{g}} \\
 \frac{S_\mathrm{t}^2}{2 f_\mathrm{g} \left(S_\mathrm{t}^2+1\right)}-\frac{\mathrm{i} \kappa_x S_\mathrm{t}^4}{2 f_\mathrm{g} \left(S_\mathrm{t}^2+1\right)^2} & 0 & -\frac{\mathrm{i} \kappa_x S_\mathrm{t}^2}{f_\mathrm{g} \left(S_\mathrm{t}^2+1\right)} & 0 & \frac{S_\mathrm{t}}{2} & -\frac{\mathrm{i} (2 f_\mathrm{p}-1) \kappa_x S_\mathrm{t}^2}{f_\mathrm{g} \left(S_\mathrm{t}^2+1\right)}-\mathrm{i} \omega  S_\mathrm{t}+1 & 0 & 0 \\
 0 & 0 & 0 & -\frac{\mathrm{i} \kappa_x S_\mathrm{t}^2}{f_\mathrm{g} \left(S_\mathrm{t}^2+1\right)} & 0 & 0 & -\frac{\mathrm{i} (2 f_\mathrm{p}-1) \kappa_x S_\mathrm{t}^2}{f_\mathrm{g} \left(S_\mathrm{t}^2+1\right)}-\mathrm{i} \omega  S_\mathrm{t}+1 & -\frac{\mathrm{i} \kappa_z S_\mathrm{t}}{f_\mathrm{g}} \\
 \frac{\kappa_x S_\mathrm{t}}{S_\mathrm{t}^2+1} & \kappa_x & 0 & \kappa_z & -f_\mathrm{p} \kappa_x & 0 & -f_\mathrm{p} \kappa_z & 0 \\
\end{vmatrix} .
\end{small}
\end{equation}
\end{multicols}

\section{3rd order polynomial}
\label{sec:app_order3}

\begin{multicols}{1}
\begin{eqnarray}
    \mathcal{P}_{8}^{(3)}(\omega) & = & \mathrm{i} \frac{S_\mathrm{t}^3}{f_\mathrm{g}} \omega^6 \nonumber \\
 &&   -\frac{2+f_\mathrm{g}}{f_\mathrm{g}} S_\mathrm{t}^2 \omega^5 \nonumber \\
&&    + \left\lbrace -\mathrm{i} \frac{1+2 f_\mathrm{g}}{f_\mathrm{g}} S_\mathrm{t} + \frac{S_\mathrm{t}^3}{f_\mathrm{g}^2}\left(-\mathrm{i}f_\mathrm{g}\left(1 + \cos^2 \theta \right) - 2 \kappa_x - 6 f_\mathrm{g}\kappa_x  + 11 f_\mathrm{g}^2 \kappa_x \right) \right\rbrace \omega^4  \nonumber \\
&&    + \left\lbrace 1 + \frac{S_\mathrm{t}^2}{f_\mathrm{g}} \left( 2 + f_\mathrm{g}(3 \cos^2 \theta -1) - 6 \mathrm{i} \kappa_x f_\mathrm{p} \right) \right\rbrace \omega^3 \nonumber \\
&&    + \left\lbrace \left( -\mathrm{i} (1 - 3 \cos^2 \theta) + \frac{\mathrm{i}}{f_\mathrm{g} } + \kappa_x \right)S_\mathrm{t} + \left( \kappa_x\left(5 -15 \cos^2 \theta + \frac{2}{f_\mathrm{g}^2} - \frac{7}{f_\mathrm{g}} + \frac{12}{f_\mathrm{g}}\cos^2 \theta + 4 \mathrm{i} \kappa_x - \frac{ \mathrm{i} \kappa_x}{f_\mathrm{g}}\right) + \frac{\mathrm{i} \cos^2 \theta}{f_\mathrm{g}}  \right) S_\mathrm{t}^3 \right\rbrace \omega^2  \nonumber \\
&&    + \cos^2 \theta \left\lbrace -1 + S_\mathrm{t}^2 \left(-1 + 3 \mathrm{i}  \kappa_x\left(\frac{1}{f_\mathrm{g}} -1 \right) \right) \right\rbrace \omega \nonumber \\
&&    +\kappa_x \cos^2 \theta \left\lbrace \left(-2 +\frac{1}{f_\mathrm{g}} \right) S_\mathrm{t} + \left( 2 - \frac{2}{f_\mathrm{g}} - 12 \mathrm{i} \kappa_x \left(1 - \frac{1}{f_\mathrm{g}} \right) - 3 \mathrm{i} \frac{\kappa_x}{f_\mathrm{g}^2}  \right) S_\mathrm{t}^3 \right\rbrace . \label{eq:apP3}
\end{eqnarray}
\end{multicols}

\section{Coefficients of the Euclidian factorisation}
\label{sec:coeffs_p3}

\begin{multicols}{1}
\begin{eqnarray}
    \alpha_1 & = & -\frac{f_\mathrm{p}}{2 f_\mathrm{g}} \left(\kappa_x + \mathrm{i} \sin^2 \theta \right),  \\
    \alpha_2 & = & \frac{1}{2 \cos \theta} \frac{f_\mathrm{p}}{f_\mathrm{g}^2}  \left( \frac{f_\mathrm{p} \kappa_x^2 }{4} - \frac{\sin^4 \theta f_\mathrm{p} }{4} - (f_\mathrm{p} - f\mathrm{g}) \kappa_x^2  \right. 
   + \cos^2 \theta \sin^2 \theta - \left. \mathrm{i} \kappa_x  \left( f_\mathrm{p}  - \frac{\sin^2 \theta}{2} (1+f_\mathrm{g}) \right) \right), \nonumber \\
    \alpha_3 & = &\frac{f_{\rm p} \kappa_x}{4 f_{\rm g}^3}\left(  (1+f_\mathrm{g}(1+f_\mathrm{g})) - \left(1-f_\mathrm{p}(1+f_\mathrm{p}) \right)\cos\left(2 \theta \right)  - 2 \frac{(f_\mathrm{p}-f_\mathrm{g})^2}{\cos^2 \theta} \kappa_x^2 \right)      - \frac{\mathrm{i} f_{\rm p}}{2 f_{\rm g}^3} \left( f_\mathrm{p}(2+f_\mathrm{g})\sin^4 \theta + f_\mathrm{g} (1+ 2 f_\mathrm{p}) \kappa_x^2 - \sin^2 \theta + \frac{(f_\mathrm{p}-f_\mathrm{g})^2}{\cos^2 \theta} \kappa_x^2 \right) , \nonumber  \\
    \beta_0 & = & - S_\mathrm{t} \frac{\kappa_x (f_\mathrm{p}-f_\mathrm{g})}{f_\mathrm{g}} 
    + S_\mathrm{t}^3 \left( -\kappa_x \left(-\frac{2 f_\mathrm{p}}{f_\mathrm{g}} + \frac{12 \mathrm{i} \kappa_x f_\mathrm{p}}{f_\mathrm{g}} - \frac{3 \mathrm{i} \kappa_x}{f_\mathrm{g}^2} \right) \right. 
    \left. + \frac{\kappa_x (f_\mathrm{p} - f_\mathrm{g})}{f_\mathrm{g}} \left( \frac{2 \alpha_2}{\cos \theta} - \frac{\alpha_1^2}{\cos^2 \theta} \right) \right) , \nonumber \\
    \beta_1  & = & 1 + \frac{S_\mathrm{t}^2}{f_\mathrm{g}} \left(  f_\mathrm{g} - \frac{f_\mathrm{p}}{f_\mathrm{g}}\sin^2 \theta - \mathrm{i} \frac{\kappa_x f_\mathrm{p}}{f_\mathrm{g}} \left( 4f_\mathrm{g} -1 \right)\right), \nonumber\\
    \beta_2  & = & -\mathrm{i} \frac{1+ 2 f_\mathrm{g}}{f_\mathrm{g}} S_\mathrm{t} 
    + \frac{S_\mathrm{t}^3}{f_\mathrm{g}^2} \left( -\mathrm{i} f_\mathrm{g} + \mathrm{i} f_\mathrm{p} \left(2 + f_\mathrm{g} \right) \sin^2 \theta \right. 
    \left. - 7 f_\mathrm{p} f_\mathrm{g} \kappa_x + 3 f_\mathrm{g}^2 \kappa_x \right) . \nonumber
\end{eqnarray}
\end{multicols}

\end{document}